\documentclass[%
 reprint,
 superscriptaddress,
 amsmath,amssymb,
 aps,
pra,
]{revtex4-2}

\usepackage{graphicx}
\usepackage{dcolumn}
\usepackage{bm}
\usepackage{hyperref}

\hypersetup{
    unicode=false,          
    pdftoolbar=true,        
    pdfmenubar=true,        
    pdffitwindow=false,     
    pdfstartview={FitH},    
    pdftitle={Modelling spatio-temporal dynamics of chiral coupling of quantum emitters to light fields in nanophotonic structures},    
    pdfauthor={Olthaus, Sohr, Wong, Oh, Reiter},     
    pdfsubject={},   
    pdfcreator={},   
    pdfproducer={}, 
    pdfkeywords={} {} {}, 
    pdfnewwindow=true,      
    colorlinks=true,       
    linkcolor=blue, 
    citecolor=blue,        
    filecolor=blue,      
    urlcolor=blue,           
	breaklinks=true
}

\begin{document}

\preprint{APS/123-QED}

\title{Modelling spatio-temporal dynamics of chiral coupling of quantum emitters to light fields in nanophotonic structures}

\author{Jan Olthaus}
 \affiliation{Institut f\"ur Festk\"orpertheorie, Universit\"at M\"unster, 48149 M\"unster, Germany}
 
\author{Maximilian Sohr}
\affiliation{Institut f\"ur Festk\"orpertheorie, Universit\"at M\"unster, 48149 M\"unster, Germany}

\author{Stephan Wong}
 \affiliation{School of Physics and Astronomy, Cardiff University, Cardiff CF24 3AA, United Kingdom}

\author{Sang Soon Oh}
 \affiliation{School of Physics and Astronomy, Cardiff University, Cardiff CF24 3AA, United Kingdom}
 
\author{Doris E. Reiter}
 \email{doris.reiter@uni-muenster.de}
\affiliation{Institut f\"ur Festk\"orpertheorie, Universit\"at M\"unster, 48149 M\"unster, Germany}
\affiliation{Condensed Matter Theory, Department of Physics, TU Dortmund, 44221 Dortmund, Germany}

\date{\today}

\begin{abstract}
A quantum emitter placed in a nanophotonic structure can result in non-reciprocal phenomena like chiral light excitation. Here, we present a theoretical model to couple circularly polarized emitters described by the density matrix formalism to the electromagnetic fields within a finite-difference time-domain (FDTD) simulation. In particular, we discuss how to implement complex electric fields in the simulation to make use of the rotating wave approximation. By applying our model to a quantum emitter in a dielectric waveguide and an optical circulator, we show how the excitation of the quantum system depends on its position and polarization. In turn, the backcoupling can result in strongly asymmetric light excitation. Our framework and results will help better understand spatio-temporal dynamics of light field in nanophotonic structures containing quantum emitters.

\end{abstract}

\maketitle

\section{\label{sec:intro}Introduction}

Chiral quantum optics presently is one of the main contributors to advancements in the fields of nanophotonics, quantum optics and quantum information processing \cite{chiral_rev}. 
Chirality in these systems generally refers to the possibility to link the local light polarization of electromagnetic sources to the direction of light propagation in a nanophotonic structure. The underlying nanophotonic structure can vary depending on the specific applications, ranging from simple dielectric waveguides \cite{chiral_cpl_wvg} to nanopatterned structures \cite{perov_chiral}, line defects in \(2\)D photonic crystals \cite{chiral_cpl_2dphc, chiral_phc}, \(2\)D photonic crystal slabs \cite{2d_ga}, or topological structures \cite{topo_2d, topo_chiral_2dphc, topo_st,add-drop}. Whereas light propagation in passive and linear nanophotonic devices is fully reciprocal, chiral coupling with quantum emitters is a non-reciprocal process as it involves temporal dynamics showing nonlinear optical responses \cite{chiral_rev,nreci}. The realization of such non-reciprocal devices is based on integrated single photon emitters \cite{schrinner_nano,eich,hoese,coupling_JO} and can be used to develop technologies like spin-quantum entanglement/spin transfer \cite{chiral_cpl_wvg2,Pucher2022}, isolators \cite{appl_iso}  and quantum optical circulators \cite{qo_circ}. All of these systems require strong light-matter interactions via light confinement and can be used as building blocks for quantum information processes.

 There are different available approaches to model the interaction between emitters and nanophotonic systems like the Green's function approach \cite{em_array}, or the input-output formalism \cite{chiral_cpl_wvg}. However, these models have limitations in studying spatio-temporal dynamics of the light matter interaction because both approaches depend on the harmonic solutions obtained by solving frequency-domain Maxwell's equations.
 On the other hand, the density matrix formalism is a well-understood tool and has been commonly used to model multi-level systems \cite{2lvl_uni,3lvl_uni}. Indeed, the density matrix formalism can be used to describe the emitters with transitions sensitive to circularly polarized light \cite{3lvl_complex_1dfdtd}. 
 
 In this paper, we develop a model to analyze the dynamics of the chiral coupling of a single emitter to nanophotonic structures. To study the coupling of the emitter to the light fields in nanophotonic structures, we incorporate the density matrix formalism into the simulation of the classical electromagnetic fields using a 2D finite-difference time-domain (FDTD) solver. This method enables us to analyze all temporal dynamics within the device. Another advantage of our model is that we account for circularly polarized coupling in 2D FDTD, surpassing state-of-the-art implementations \cite{2lvl_kevin,2lvl_uni_2DFDTD,3lvl_complex_1dfdtd}. Further, we run the FDTD simulations using complex fields, such that we can apply the rotating-wave approximation (RWA) to increase the stability of interactions between the emitter and classical fields.

Our numerical approach with 2D FDTD also allows us to consider more complex structures. To demonstrate chiral coupling with the developed model, we apply it to two examples. First, we consider a quantum system composed of a dielectric waveguide and a single quantum emitter. Here, we show that we can selectively excite unidirectional waveguide modes with the backcoupling fields induced by the emitter depending on its position. Then, we consider an optical circulator coupled to an emitter \cite{qo_circ, bot_mr, em_res, om_circ, om_circ2, qoc_exp, qoc_2}, which is a promising structure for quantum information technology. We show that our model can capture the temporal dynamics of both the emitter's population and the optical fields within the system. 

\section{Theoretical model}
The main focus of our approach is to develop a model to account for chiral coupling between the propagating electromagnetic field and a quantum few-level system. In this section, we introduce the Stokes parameters to describe the polarization of the local field of the mode and the density matrix formalism for the light-matter interaction.

\subsection{Chirality}
The polarization of the local light field $\bm{E} = (E_x, E_y)$ in a 2D system can be encoded in the Stokes parameters \cite{stokes_par,collett}
\begin{subequations}
\begin{eqnarray}
S_0 &=& |E_{\text{x}}|^2 + |E_{\text{y}}|^2\\
S_1 &=& \frac{|E_{\text{x}}|^2 - |E_{\text{y}}|^2}{S_0}\\
S_2 &=& \frac{2\cdot \text{Re} \left(E_{\text{x}}^*\cdot E_{\text{y}}\right)}{S_0}\\
S_3 &=& \frac{2\cdot \text{Im} \left(E_{\text{x}}^*\cdot E_{\text{y}}\right)}{S_0}\,.
\end{eqnarray}
\end{subequations}
While $S_0$ is a measure of the intensity of the light, the polarization is encoded in $S_1, S_2$ and $S_3$. The chirality of light can be deduced from $S_3$, where $S_3=\pm 1$ corresponds to left/right circularly polarized light and $S_3=0$ indicates linearly polarized light. When analyzing chirality in nanophotonic structures, it is of high importance to identify locations where the light has a pure circular polarization. Those locations are called C-points \cite{c_point}.

\subsection{\label{sec:model}Coupling to quantum emitter}
Next, we want to couple the light to a quantum emitter, whose dipole transitions are sensitive to circularly polarized light. As an example we look at a circularly polarized \(2\)-level system. An extrapolation to more complex systems is straightforward. The Hamiltonian $\cal H$ of the \(2\)-level system coupling to a complex electric field $\bm E$ and dipole moment $\bm \mu$ using RWA reads
\begin{eqnarray}
\arraycolsep=2.0pt\def\arraystretch{1.5}
{\cal{H}} = {\cal{H}}_0 - \bm{\mu} \cdot \bm{E} = \left( \begin{array}{cc} 0 & -\bm{\mu}_{12} \bm{E} \\ -\bm{\mu}_{12}^* \bm{E}^* & \hbar \omega_2 \\\end{array}\right)\,.
\end{eqnarray}
Here ${\cal{H}}_0$ is the system Hamiltonian of the two energy levels $|1\rangle$ and $|2\rangle$ separated by \(\hbar \omega_2\). The coupling is mediated by the complex transition dipole moment \(\bm{\mu}_{12}\). We choose \(\bm{\mu}_{12} = \bm{\mu}^{\text{Re}} + i\bm{\mu}^{\text{Im}}\).
The optical Maxwell-Bloch equations for the density matrix (\(\rho\)) elements are obtained by solving the Liouville's equation 
\begin{eqnarray}
i\hbar\frac{\partial}{\partial t} \rho = {\rho} \cdot {\cal{H}} - {\cal{H}} \cdot \rho \label{eq:lv}
\end{eqnarray}
Including a dephasing rate \(\gamma\), this results in:
\begin{subequations}
\begin{eqnarray}
i\hbar\frac{\partial}{\partial t} \rho_{11} &=& -\bm{\mu}_{12}^* \cdot \bm{E}^*\, \rho_{12} + \bm{\mu}_{12} \cdot \bm{E} \, \rho_{12}^* \,,\\
i\hbar\frac{\partial}{\partial t}\rho_{12} &=& -\bm{\mu}_{12} \cdot \bm{E} \, (2 \rho_{11} - 1) + \hbar \omega_2 \, \rho_{12} \\
& & - i \hbar\,\gamma\,\rho_{12}.
\end{eqnarray}
\end{subequations}
To analyze chiral coupling, we exclusively look at the coupling between the quantum emitter and transverse-electric (TE) modes of the nanophotonic systems. We set up our structures in a way that the TE-mode \(\bm{E}\)-field profiles are polarized in the (\(\bm{e}_{\text{x}},\bm{e}_{\text{y}}\))-plane. To achieve optimal conditions for non-reciprocal coupling, the quantum emitter has to be polarized in this plane. Accordingly, for all simulations we choose \(\bm{\mu}^{\text{Re}} ||\, \bm{e}_{\text{y}}\) and \(\bm{\mu}^{\text{Im}} ||\, \bm{e}_{\text{x}}\). We take the transition matrix element to have a circular polarization with  \(\bm{\mu}_{12} = \mu \cdot \bm{e}_{\text{y}} \pm i \mu \cdot \bm{e}_{\text{x}}\), as well as \(|\bm{\mu}_{12}^{\text{Re}}| = |\bm{\mu}_{12}^{\text{Im}}| = \mu\), yielding either a left-handed (\(\circlearrowleft\), ``+") or a right-handed (\(\circlearrowright\), ``-") circularly polarized system.
Using \(\bm{E} = \bm{E}^{\text{Re}} + i \bm{E}^{\text{Im}}\), we can rewrite the dynamical equations to the optical Bloch equations
\begin{subequations} \label{eq:eom_rho}
\begin{eqnarray}
\frac{\partial}{\partial t}\rho_{11} &=& 
    \frac{2}{\hbar}\mu\left[ \left(E_{\text{y}}^{\text{Im}} \pm E_{\text{x}}^{\text{Re}}\right)\rho_{12}^{\text{Re}} 
    - \left( E_{\text{y}}^{\text{Re}}\mp E_{\text{x}}^{\text{Im}}\right)\rho_{12}^{\text{Im}} \right]\hspace{5mm}
    \label{eq:eom_11}\\
\frac{\partial}{\partial t}\rho_{12}^{\text{Im}} &=& 
    \frac{1}{\hbar}\mu\left( E_{\text{y}}^{\text{Re}} \mp  E_{\text{x}}^{\text{Im}}\right) \cdot \left(2 \rho_{11} - 1\right) - \omega_2 \rho_{12}^{\text{Re}} \nonumber\\
    & & -\gamma\rho_{12}^{\text{Im}}\,
    \label{eq:eom_Re12}\\
\frac{\partial}{\partial t}\rho_{12}^{\text{Re}} &=& 
    - \frac{1}{\hbar}\mu \left( E_{\text{y}}^{\text{Im}} \pm  E_{\text{x}}^{\text{Re}}\right) \cdot \left(2 \rho_{11} - 1\right) + \omega_2 \rho_{12}^{\text{Im}} \nonumber\\
    & & - \gamma \rho_{12}^{\text{Re}}
    \label{eq:eom_Im12} .
\end{eqnarray}
\end{subequations}
So far, we did not make any assumption about the local field polarization of the TE-mode profiles at the position of the quantum emitter. If the local field polarization is perfectly circular, we are in a C$^{(\circlearrowleft / \circlearrowright)}$-point. In a C-point with right-handed circular polarization, we then have $E_y^{\text{Im}}=E_x^{\text{Re}}$ and $E_y^{\text{Re}}= -E_x^{\text{Im}}$, showing that both terms in Eq.~\eqref{eq:eom_11} either enhance or cancel each other depending on the polarization of the quantum emitter. If we switch to a C-point with left-handed circular polarization, we have $E_y^{\text{Im}} = -E_x^{\text{Re}}$ and $E_y^{\text{Re}}= E_x^{\text{Im}}$. As can be seen the canceled and enhanced cases interchange compared to the first case, demonstrating the sensitivity to the chirality of coupling between field and emitter. 

The optical Bloch equations in Eq.~\eqref{eq:eom_rho} describe the effect of the classical fields on the dynamics of the emitter population. In this paragraph, we describe the coupling of the emitter back into the field. For this, we make use of the macroscopic polarization induced by the interaction \cite{3lvl_complex_1dfdtd}
\begin{eqnarray} \label{eq:pol}
\bm{P}(t) &=& 2\left[ \bm{\mu}_{12}^*\rho_{12} \right] =  2 \mu \left[\left(\bm{e}_{\text{y}} \mp i \, \bm{e}_{\text{x}}\right)\rho_{12} \right]\,\,\,\,\,\,\,\, \notag \\
&=& 2 \mu \left[\left(\rho_{12}^{\text{Re}} + i \rho_{12}^{\text{Im}} \right)  \bm{e}_{\text{y}} \mp \left( i \rho_{12}^{\text{Re}} - \rho_{12}^{\text{Im}}\right)  \bm{e}_{\text{x}} \right]\,.
\end{eqnarray}
This polarization is then coupled back into the classical Maxwell equations at the position of the emitter.

The extension of this model to a multi-level emitter is straightforward. In the results section we will consider a V-type 3-level system as schematically shown in Fig.~\ref{fig:wvg_no_bcpl}(a) consisting of the ground state $|1\rangle$ and two excited states $|2\rangle = |\text{R}\rangle$ and $|3\rangle = |\text{L}\rangle$. We assume that the dipole moment for the transition $1 \longleftrightarrow \text{R}$ is given by \(\bm{\mu}_{1R} = \bm{\mu}^{\text{Re}} - i\bm{\mu}^{\text{Im}}\) (\(\circlearrowright\), ``-") and for the transition $1 \longleftrightarrow \text{L}$ is \(\bm{\mu}_{1L} = \bm{\mu}^{\text{Re}} + i\bm{\mu}^{\text{Im}}\) (\(\circlearrowleft\), ``+"). To accurately describe the V-type character, we prevent transitions between the excited states \(\bm{\mu}_{RL} = \bm{\mu}_{LR} = 0\). Furthermore, both excited states have the same energy \(\omega_L = \omega_R = \omega_2\). 

\subsection{Coupling to electromagnetic fields}

To describe the time evolution of the electric and magnetic fields, we  solve the Maxwell's equations using the FDTD method and our solver is based on Ref. \cite{fdtd_schneider}. This method is based on the discretization of time and space into a grid. The spatial resolution of our simulations is determined by the quality of the gridpoints closest to the C-points of our nanophotonic structures. We find that for resolutions higher than \(\Delta x = \lambda_2/240\), the polarization in the gridpoints is close enough to circular, so that the effects of the deviation on the emitter populations can be neglected. All results presented in this paper are based on simulations with resolution \(\Delta x = 3\,\text{nm}\) and timestep \(\Delta t = 7.25\,\text{ns}\). As the absorbing boundary condition, we implemented a split-field perfectly matched layer.
Note that different to most FDTD implementations, we work with complex fields to make use of the RWA in the Bloch equations. This increases the stability of the numerical integration of the few-level system significantly. 
At the gridpoint of the emitter position, the update equations for the electric field components include the effects of the macroscopic polarization induced by the emitter, which has been defined in the previous section. Here we restrict our simulations to the 2D TE-case. As such the emitter placed in a grid point corresponds to a line of emitters in the perpendicular direction due to the periodic boundary conditions. If one would extent the model to 3D, the emitter-field coupling would be significantly smaller. Furthermore losses would be introduced due do out-of-plane field scattering of the waveguide mode as well as the backcoupling fields. However, despite these quantitative changes the qualitative results are not affected by the restriction to 2D. We set up the simulation cell so that the electric field lies within the (\(\bm{e}_{\text{x}},\bm{e}_{\text{y}}\))-plane and the magnetic field is aligned with the \(\bm{e}_{\text{z}}\)-direction. The update equations at the point of the emitter are:
\begin{subequations}\begin{eqnarray}
\frac{\partial E_{\text{x}}}{\partial t} &=& \frac{1}{\epsilon}\frac{\partial H_{\text{z}}}{\partial y} - \frac{\sigma}{\epsilon} E_{\text{x}} - \frac{1}{\epsilon}\frac{\partial}{\partial t} P_{\text{x}} , \\
\frac{\partial E_{\text{y}}}{\partial t} &=& -\frac{1}{\epsilon}\frac{\partial H_{\text{z}}}{\partial x} - \frac{\sigma}{\epsilon} E_{\text{y}} - \frac{1}{\epsilon}\frac{\partial}{\partial t} P_{\text{y}}  , \\
\frac{\partial H_{\text{z}}}{\partial t} &=& \frac{1}{\mu}\left( \frac{\partial E_{\text{x}}}{\partial y} - \frac{\partial E_{\text{y}}}{\partial x}\right) + \frac{\sigma_m}{\mu} H_{\text{z}} .
\end{eqnarray}\end{subequations}
Here \(\sigma\) and \(\sigma_m\) are the electric and magnetic conductivities, respectively (here: \(\sigma = \sigma_m = 0\)). 

\section{\label{sec:model_intro} Quantum emitter in a dielectric waveguide}

Using the theoretical description of the quantum emitter, we now embed it into a nanophotonic structure, which we model within a FDTD simulation. Inspired by the solid-state quantum emitters, we set the transition energy of our quantum emitter to \(\lambda_2 = 738\,\text{nm} \left(\omega_2 = 406\,\text{THz}\right)\) \cite{2d_ga}. As a first system, we consider a dielectric waveguide to characterize the chiral coupling within our model.

\subsection{\label{sec:wvg} System Characterization}
\begin{figure}[t]
\includegraphics[width=\columnwidth]{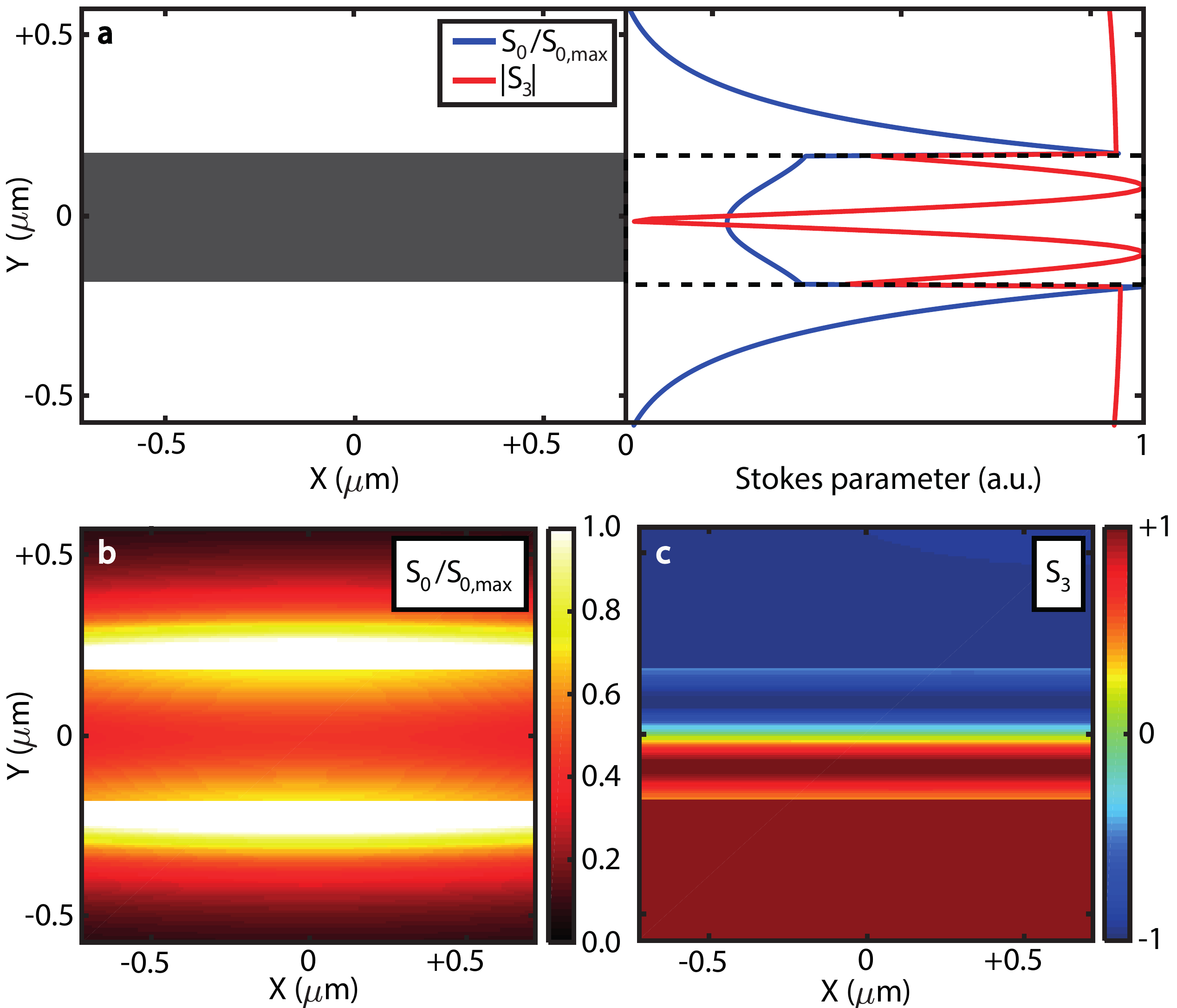}
\caption{\label{fig:chiral_wvg} 
(a) Left: Schematic of dielectric structure. Right: Line plot of the stokes parameters \(S_0\) and \(S_3\) through the cross section of the waveguide. Color plot of the Stokes parameters (b) \(S_0\) and (c) \(S_3\).}
\end{figure}

We start by characterizing the waveguide without the emitter to identify possible C-points. Based on the transition energy \(\omega_2\) of the single emitter, we choose a waveguide width of \(b=157\,\text{nm}\) and permittivity \(\epsilon = 6\). The waveguide is designed based on simulations using the iterative eigen-frequency solver MPB \cite{MPB_tut} to ensure it only supports the fundamental \(\text{TE}_{10}\) mode at the resonant frequency. Because we restrict ourselves to 2D simulations, this results in a relatively small waveguide width. Figure~\ref{fig:chiral_wvg} shows the relevant Stokes parameters of the fundamental TE-mode after resonant excitation with \(S_0\) in Fig.~\ref{fig:chiral_wvg}(b) and \(S_3\)  in Fig.~\ref{fig:chiral_wvg}(c). We remind that the first Stokes parameter (\(S_0\)) gives the intensity, which in turn determines the strength of the light-matter interaction, while \(S_3 = \pm 1\) (also: \(S_1 = S_2 = 0\)) denotes C-points. Note that the sign of \(S_3\) also depends on the propagation direction of the excited mode. In Fig.~\ref{fig:chiral_wvg}, we study a mode propagating from the left (\(x<0\)) to the right (\(x>0\)). If we were to look at the opposite case  (from right (\(x>0\)) to left (\(x<0\))), the sign of all \(S_3\) values would switch. 
Because of the small waveguide width, the \(E_x\) field component of the mode profile is large compared to the \(E_y\) field. This results in a local minimum of the total electric field strength \(S_0\) in the center of the waveguide. The field strength increases with distance from the center. Additionally, there are large spikes of the evanescent electric fields at the waveguide-air interfaces. In the center of the waveguide, the mode is linearly polarized ($S_3 = 0$, $\bm{E} \parallel \bm{e}_{\text{y}}$). When increasing the distance from the center, the mode becomes elliptically polarized and shows C-points close to the waveguide boundaries (at \(y = \pm 43\,\text{nm}\)). The evanescent fields close to the waveguide are almost circularly polarized. The best positions to achieve chiral coupling are those that combine high \(S_0\) values with \(S_3\) values close to \(\pm1\). The best option in this case is emitter placement along the line of C-points in the waveguide material. A good secondary option would be placement of the emitter in the evanescent fields.

\subsection{\label{sec:wvg_no_bcpl} Excitation of the quantum emitter}

\begin{figure}[t]
\includegraphics[width=\columnwidth]{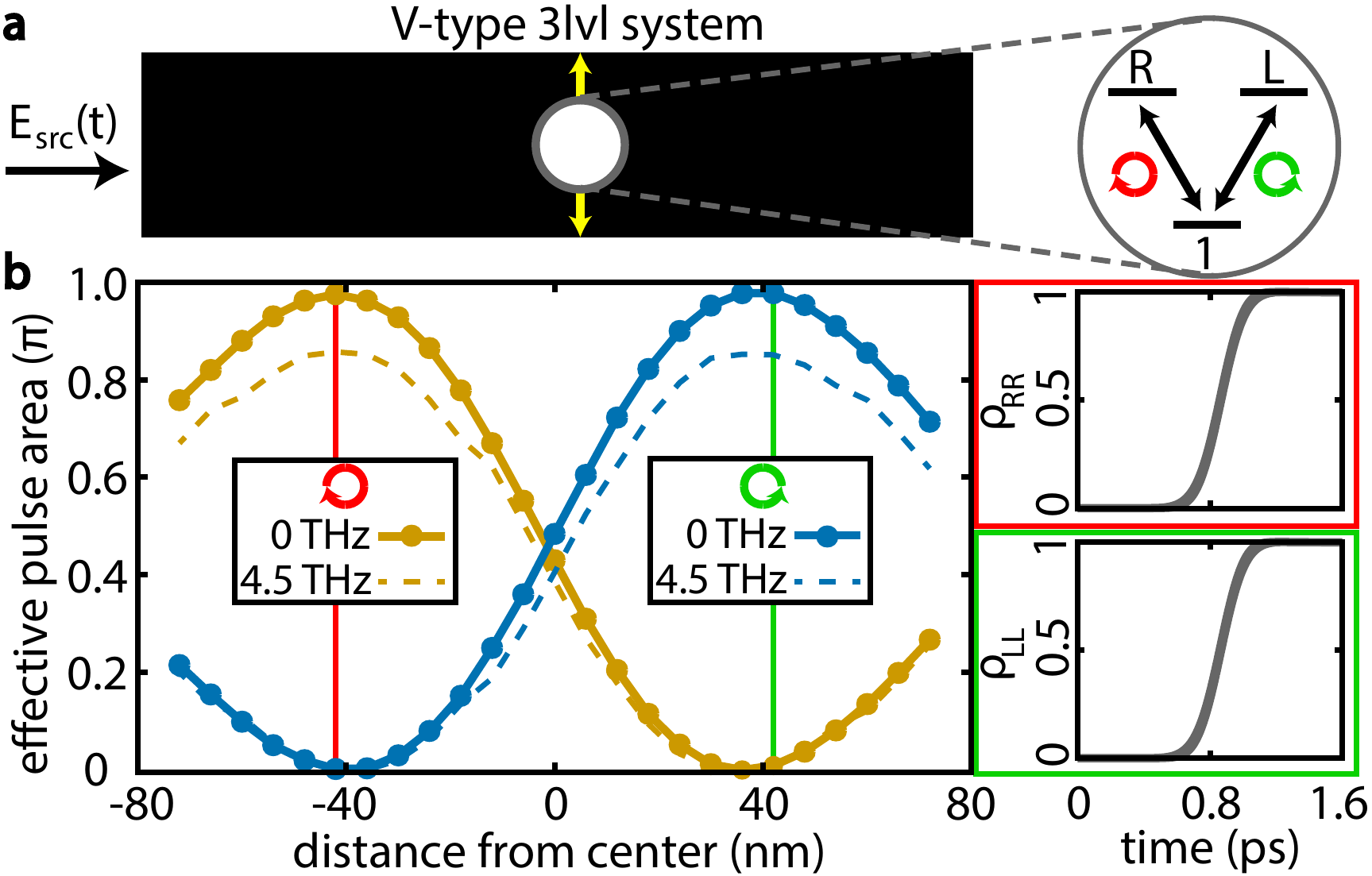}
\caption{\label{fig:wvg_no_bcpl} (a) Sketch of the  waveguide with a quantum emitter with a V-type three-level structure as inset excited by an external source. (b) Left: effective pulse area for two emitter polarizations as a function of the distance from the waveguide center. Included are results for the model without dephasing (\(\gamma = 0\,\text{THz}\)) and with dephasing (\(\gamma = 4.5\,\text{THz}\)). Right: population dynamics of the density matrix for the two states $|\text{L}\rangle$ and $|\text{R}\rangle$ in the C-points without dephasing.}
\end{figure}

\begin{figure*}[ht]
\includegraphics[width=\textwidth]{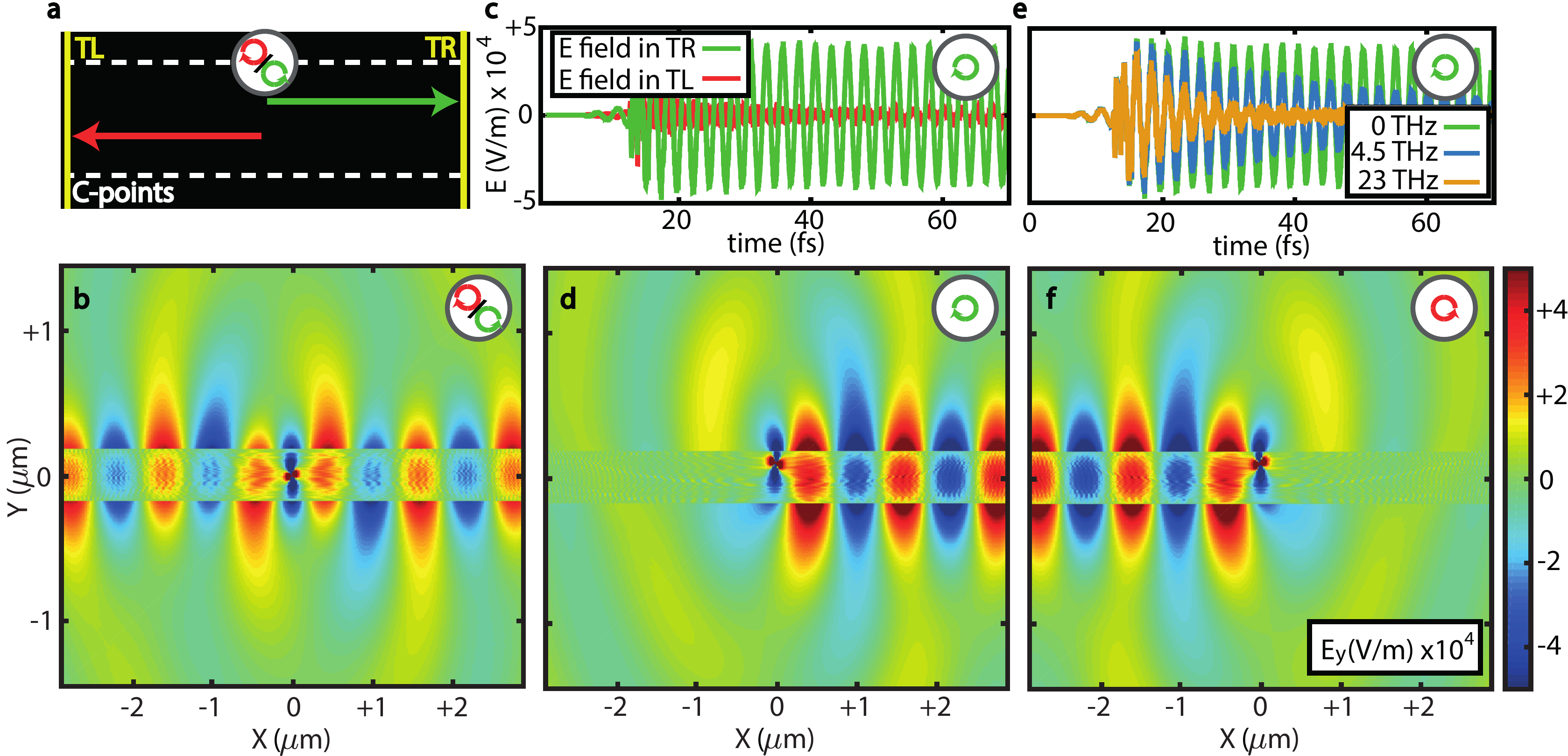}
\caption{\label{fig:wvg_bcpl} (a) Sketch of the waveguide with an excited quantum emitter. The electric field introduced by the emitter backcoupling propagating through two planes on the left (TL) and right (TR) side of the grid is monitored.  (b),(d),(f) Snapshots of the electric field  for the emitter placed in the center (b) for either occupation, the upper C-point for \(\rho_{LL} = \frac{1}{2}\) (d) and for \(\rho_{RR} = \frac{1}{2}\) (f).  Monitored fields in TL/TR plane for the emitter in the upper C-point for \(\rho_{LL} = \frac{1}{2}\) in TL/TR plane for \(\gamma=0\,\text{THz}\) (c) and in TR plane for different dephasing rates \(\gamma=0\,\text{THz},\, 4.5\,\text{THz}\) and \(23\,\text{THz}\) (e).}
\end{figure*}

Now, we couple the density matrix approach model to the classical 2D-FDTD simulations. We add the quantum emitter to our simulations as sketched in Fig.~\ref{fig:wvg_no_bcpl}(a) and consider the change of population induced by the field. For now, we exclusively want to focus on the excitation of the emitter. In other words, we neglect the backcoupling induced by the macroscopic polarization [Eq.~\eqref{eq:pol}].

Using the FDTD simulation, we calculate the fields in the waveguide, which are induced by an external electric field \(\left(\bm{E}_{\text{src}}\left(t\right)\right)\) placed on the left side of the cell in the center of the waveguide. For the source we assume a Gaussian point-dipole source given by
\begin{eqnarray}
\bm{E}_{\text{src}}\left(t\right) = e^{\frac{-\left(t - t_{0}\right)^2}{2 \text{w}^2}}\, e^{-i\omega_2 \left(t - t_0\right)}\, \bm{e}_{\text{y}}\,.&&
\label{eq:Esrc}
\end{eqnarray}
The quantum emitter is placed in the middle of the cell with a variation along the y-axis as indicated by the yellow line in Fig.~\ref{fig:wvg_no_bcpl}~(a). The initial emitter populations are \(\rho_{11} = 1\) and \(\rho_{\text{RR}/\text{LL}} = 0\). Depending on the distance from the waveguide center, the local electric field polarization will vary between linearly and circularly polarized. 

We then calculate the population of the quantum emitter as a function of time, for different emitter positions along the y-axis. Fig.~\ref{fig:wvg_no_bcpl}(b) shows the total change of population of the 3-level system for both emitter polarizations. As a measure we use the effective pulse area, where $\pi$ corresponds to a full excitation of the system after the pulse. We show results based on the model without dephasing (\(\gamma = 0\,\text{THz}\)) and with dephasing (\(\gamma = 4.5\,\text{THz}\)).
First the emitter is placed in the center of the waveguide. The local electric field polarization at this position is linear \(\left(\bm{E} \parallel \bm{e}_{\text{y}}\right)\). Because the transition dipole moment in this orientation is equal for both polarizations \(\left(\bm{\rho}_{1\text{R}} \cdot \bm{e}_{\text{y}} = \bm{\rho}_{1\text{L}} \cdot \bm{e}_{\text{y}} = \mu \right)\), the excitation is equal for both transitions and depends solely on the relative field strength \(S_0\). As soon as the emitter is positioned off-center, the field polarization becomes elliptical and we see chiral coupling, i.e., the effective pulse area for both transitions is distinct. The non-reciprocity is a result from the coupling to the \(E_x\) field component (see Eq.~\eqref{eq:eom_rho}). Depending on the polarization, this coupling results in either enhancement or reduction of the effective pulse area. The preferred and suppressed polarization switches for placement above/below the center. At the C-point of the waveguide mode, the enhanced case has an effective pulse area of \(A_{\text{enh}}\approx \pi\) and the suppressed case \(A_{\text{sup}}\approx 0\) in the case without dephasing. The corresponding population dynamics are shown in Fig.~\ref{fig:wvg_no_bcpl}(b), where we see that the population of the excited state rises up to one. If we compare this to the results with a dephasing rate of \(\gamma = 4.5\,\text{THz}\), we see a overall smaller coupling strength. This is most noticeable for large pulse areas, where the reduction compared to the case without dephasing becomes particularly large. Due to the limited spatial resolution of the FDTD simulations, we are not able to simulate gridpoints with perfect circular polarization. The deviation results in a small remaining excitation strength ($<1\%$) of the suppressed transition.

\subsection{\label{sec:wvg_with_bcpl} Emission behaviour}
We now include backcoupling of the emitter into the nanophotonic system. Generally, the backcoupling for single emitters creates significantly smaller fields compared to the external sources used for the excitation process of the emitter. To isolate the effect of the fields induced by the macroscopic polarization at the emitter position, we consider the emitter to have an initial occupation. This way, the emitter can emit into the nanophotonic structure without requiring external source fields.

We consider two different initial occupations for our 3-level system. In the \(\circlearrowright\)-case we assume that \(\rho_{\text{RR}} = \rho_{\text{11}} = \frac{1}{2},\, \rho_{\text{LL}} = 0\) with the only non-vanishing polarization  \(\rho_{1\text{R}}^{\text{Im}} = \frac{1}{2}\). and in the \(\circlearrowleft\)-case exchange $R$ with $L$. Because of the coupling nature, we chose a finite polarization to start the emission process according to Eq.~\eqref{eq:pol}. For small backcoupling strengths (no self-interaction/change of the initial population \cite{p_tfsf}) this results in the emitter creating oscillating fields with constant amplitudes. 

The setup of our simulations is shown in Fig.~\ref{fig:wvg_bcpl}(a). The occupied emitter is placed either in the waveguide center [Fig.~\ref{fig:wvg_bcpl}(b)] or in the upper C-point [Figs.~\ref{fig:wvg_bcpl}(c)-(f)]. We show snapshots of the electric field in Figs.~\ref{fig:wvg_bcpl}(b),(d),(f) and monitor the fields propagating through two planes on the left (TL) and right (TR) side of the simulation cell in Figs.~\ref{fig:wvg_bcpl}(c) and (e).

In the symmetric case, i.e., when an emitter is placed in the center of the waveguide, the coupling is reciprocal. Accordingly, the field is emitted in all direction, which in the waveguide geometry means it propagates right and left as indicated in the snapshot in Fig.~\ref{fig:wvg_bcpl}(b). This can be traced back to the local polarization at the waveguide center, which is linear with $S_3=0$. The emitted fields are independent of the initial occupation of the emitter. This is equivalent to the behaviour of a uniform or unidirectional emitter (reciprocal for all positions).

Next we put the emitter at a C-point in the upper half of the system, where we expect chiral coupling. Taking the initial condition of \(\circlearrowleft\)-case, i.e., only $\rho_{LL}$ is occupied, we observe a clear unidirectional emission towards the R-plane as shown in the snapshot of the electric field snapshot displayed in Fig.~\ref{fig:wvg_bcpl}(d). The directionality is validated by the monitored fields through the TL- and TR planes in Fig.~\ref{fig:wvg_bcpl}(c), which show significantly higher amplitudes in the TR-plane.     
If we now choose the other state of the 3-level system, namely the $R$-state to be occupied, we observe unidirectional emission towards the TL-plane [Fig.~\ref{fig:wvg_bcpl}(f)] and see higher amplitudes of the monitored fields in the TL-plane. Finally, we show monitored fields through the TR-plane if the L-state is initially occupied for different dephasing rates \(\gamma=0\,\text{THz},\, 4.5\,\text{THz}\) and \(23\,\text{THz}\). We can clearly observe a decay of the electric field, whose speed increases with \(\gamma\).
Note that when we place the emitter in the C-point below the waveguide center, the results of both cases interchange.  
This shows that our model is clearly capable of describing the non-reciprocal coupling in both excitation and emission behaviour of a quantum system.

\section{\label{sec:model_appl}Quantum emitter in an optical circulator}

\begin{figure*}[ht]
\includegraphics[width=\textwidth]{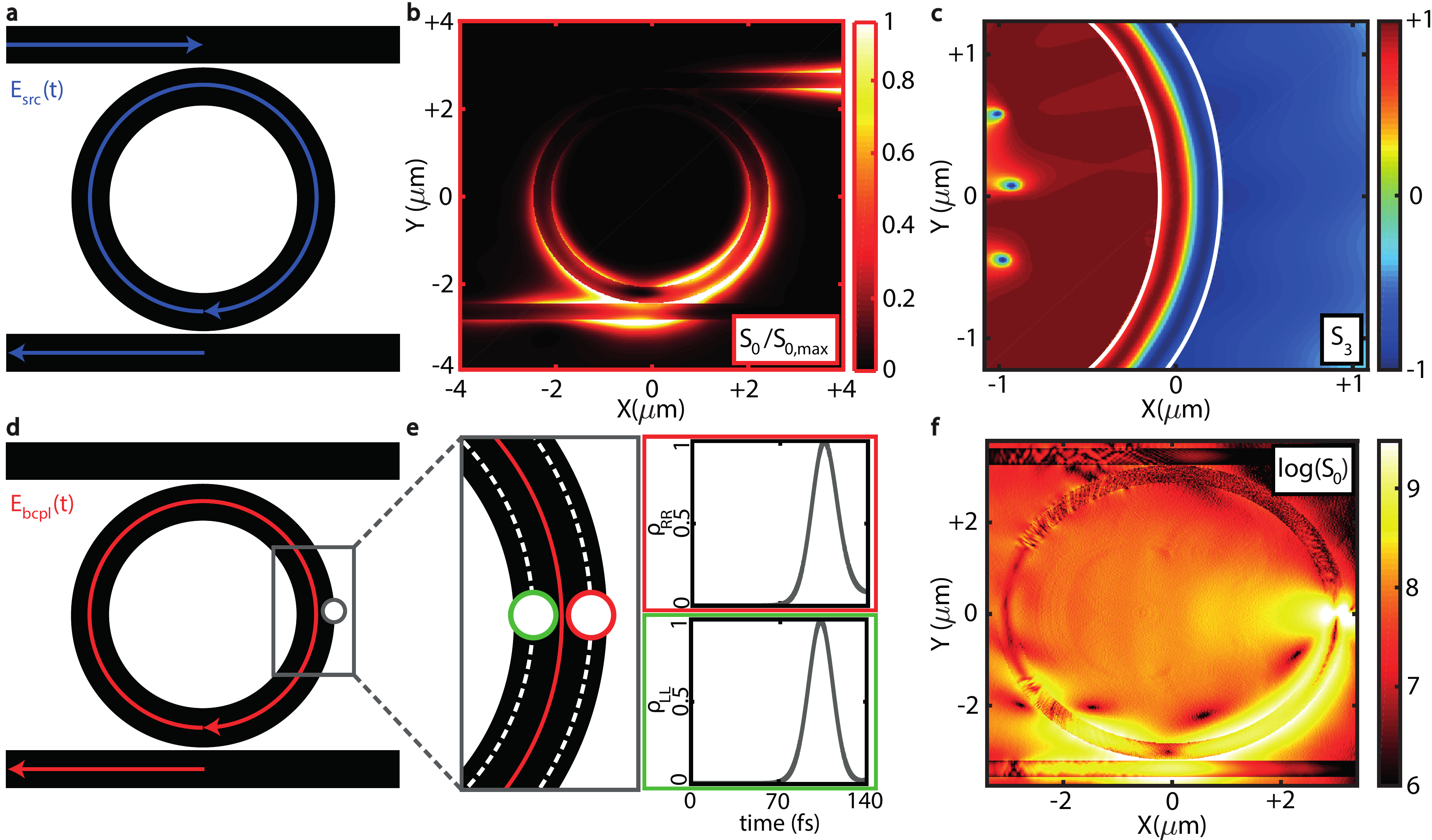}
\caption{\label{fig:circ} (a) Schematic of the structure and light propagation of the external source field \(\bm{E}_{\text{src}}\) introduced at the left side of the top waveguide of the circulator. (b) \(S_0\) snapshot of the source field. (c) Close-up of the \(S_3\) parameter of the excited ring resonator mode. The emitter is placed within the outer C-point of the ring resonator (border highlighted by white lines). (d) Schematic of the structure including emitter and propagation of the backcoupled fields induced by the emitter polarization \(\bm{E}_{\text{bcpl}}\). (e) Close-up of the ring resonator region, where the emitter is placed in one of two positions. Depending on the emitter polarization, only one state of the system will be excited (indicated by color: red \(\rightarrow\)  $|\text{R}\rangle$, green \(\rightarrow\)  $|\text{L}\rangle$). Right: population dynamics for both relevant cases. (f) Snapshot of \(\bm{E}_{\text{bcpl}}\) of the emitter in the red case.}
\end{figure*}

So far, we just considered a 1D waveguide, while our method is also capable to describe 2D patterned structures. Therefore, we apply our model to a more complex integrated non-reciprocal nanophotonic system inspired by optical circulators with a quantum emitter \cite{qo_circ, bot_mr, em_res, om_circ, om_circ2, qoc_exp, qoc_2}.
Analyzing these structures is very important and beneficial, as they can be used as scalable components for signal processing in photonic circuits and can therefore contribute to the realisation of quantum information technologies.

\subsection{\label{sec:OC_system} System characterization}

A schematic of the structure and the setup of our simulation is illustrated in Fig.~\ref{fig:circ}(a),(d). The structure consists of two dielectric waveguides in the top and bottom of the simulation cell, aligned along the \(x\) axis. These are coupled to a ring resonator in the center of the cell. The ring resonator and the waveguides have the same width and permittivity as the waveguide studied previously. Furthermore the resonators optical path length difference (OPD) is designed to support a mode resonant with the emitter transition energy \(\left(\text{OPD} = 27\lambda_2\right)\). In our simulations, we introduce electromagnetic fields using a resonant Gaussian dipole source [Eq.~\eqref{eq:Esrc}].

In Fig.~\ref{fig:circ}(a)-(c) we schematically show the propagation path of the external source field \(\bm{E}_{\text{src}}\), where Fig.~\ref{fig:circ}(b) displays a snapshot of the electric field strength \(S_0\) and Fig.~\ref{fig:circ}(c) gives the Stokes parameter \(S_3\) in a small section of the ring resonator mode. An external source (Eq.~\eqref{eq:Esrc}) excites the system at the top left arm. The field then propagates through the top waveguide from left (\(x<0\)) to right (\(x>0\)). At the waveguide-resonator interface, the clockwise-propagating mode of the ring resonator becomes excited. This mode can out-couple at the interface with the bottom waveguide, resulting in an excitation of the  waveguide mode propagating from right (\(x>0\)) to left (\(x<0\)). 

Figure~\ref{fig:circ}(c) shows $|S_3|$ over a small section of the resonator. We see that similar to the waveguide, in the middle of resonator ring the fields do not show any chirality $ |S_3|=0$, while away from the middle, we find two lines close to the air-interface where the chirality is maximal with $ |S_3|=1$, i.e., a line of C-points.
The coupling between the waveguides and the resonator occurs from the overlapping evanescent fields of both the waveguide mode and the ring resonator mode, which both have circular polarization \(\left( |S_3| \approx 1 \right)\). This results in a high degree of chirality.

\subsection{\label{sec:OC_system_RR} Coupling to quantum system}

We now place an emitter into the resonator at one of the C-points of the ring resonator mode on the right side of the resonator, as indicated by the dashed white lines in Fig.~\ref{fig:circ}(e). If we place the emitter in the C-point further to the outer edge, only the state $|\text{R}\rangle$ gets excited (red), while if we place it in the C-point further to the inner edge only $|\text{L}\rangle$ becomes occupied (green). 

This is shown in Fig.~\ref{fig:circ}(e), where we show the occupation dynamics of the two states for the two placements after excitation with the source field $\bm{E}_{\text{src}}$, as before starting without backcoupling.  While the effective pulse area in the green case is \(2\pi\), the pulse area at the outer C-point (red) is slightly smaller (\(\approx 5\%\)). The occupation of the other states stay small. The difference in occupation can be traced back to the different field strength, which is slightly larger towards the inner edge of the waveguide. If we reverse the propagation direction of the external fields, the excited states (red/green) at both emitter positions interchange. This is consistent with our previous analysis of the waveguide emitter system.

\begin{figure}[ht]
\includegraphics[width=\columnwidth]{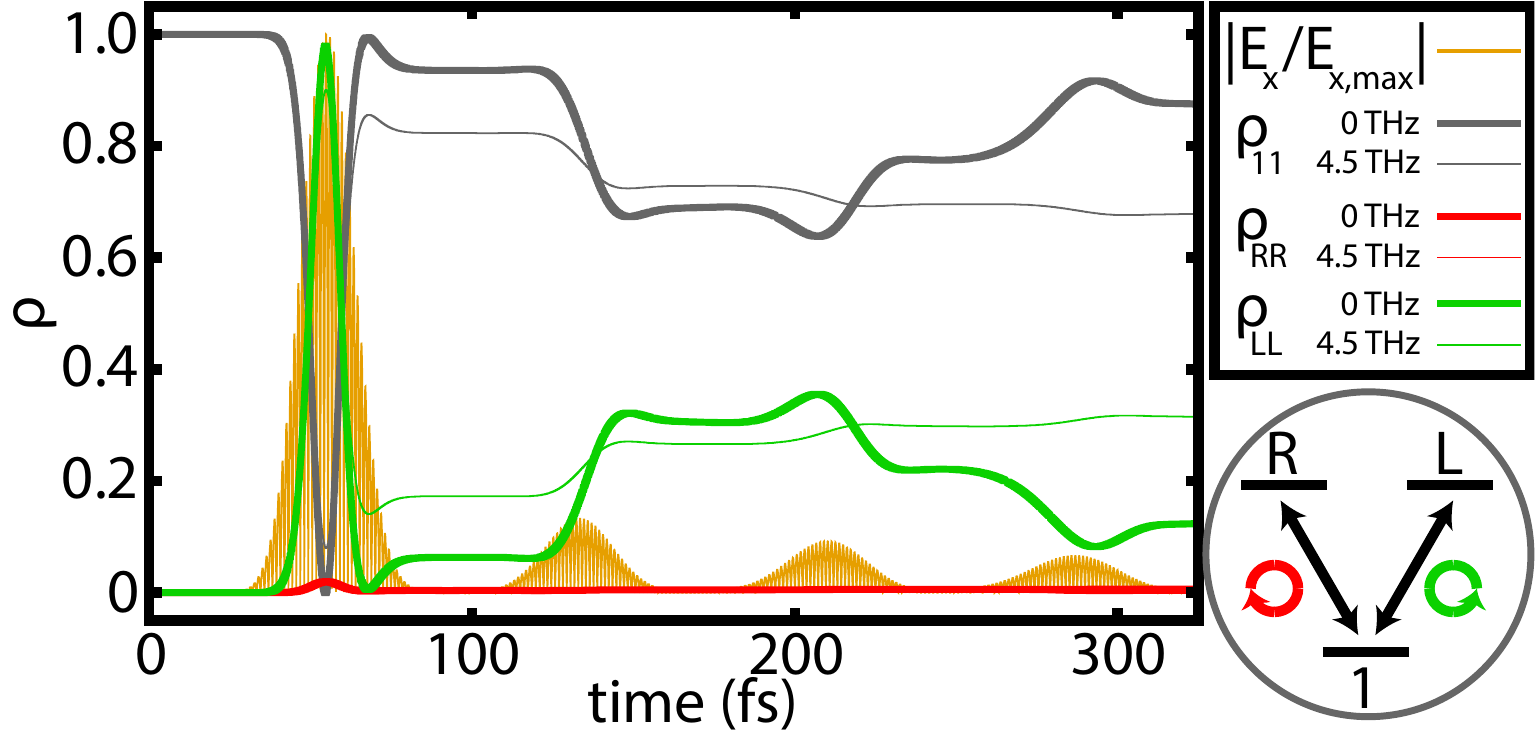}
\caption{\label{fig:rr_long} Population dynamics in the ring resonator structure after excitation with a external source field (emitter position: green). Included are results for the model without dephasing (\(\gamma = 0\,\text{THz}\)) and with dephasing (\(\gamma = 4.5\,\text{THz}\)). Additionally shown is the relative electric field strength at the emitter position \(|E_X/E_{X,\text{max}}|\). Sketch of the V-type three-level structure as inset included on the bottom right.}
\end{figure}

To demonstrate the ability of our model to capture complex time-resolved population dynamics, we show the emitter populations (emitter position: green in Fig.~\ref{fig:circ}(e)), as well as the electric fields for multiple cycles in the ring resonator, and hence multiple interactions with the quantum emitter. As can be seen in Fig.~\ref{fig:rr_long}, the initial field pulse arriving at the emitter results in a full excitation and de-excitation of the resonant L level. The small population of the non-resonant R level is a result of the slightly imperfect positioning of the emitter in the simulation grid. The pulse amplitudes decay for every subsequent pulse due to coupling to the waveguides at the top and bottom of the resonator and scattering losses. Accordingly, every subsequent pulse has a smaller pulse area. Also, some parts of the field are out-coupled each time the field passes the waveguides, which modifies the spectrum. Therefore the dynamics of the population $\rho_{LL}$ is less modified by the subsequent pulses. 

When additionally a dephasing rate of \(\gamma = 4.5\,\text{THz}\) is included, the polarization of the emitter decays within about \(200\,\text{fs}\). As we have seen in our previous analysis, this results in overall smaller coupling strengths (effective pulse areas) of the emitter compared to the case without dephasing. This is already visible when looking at the population dynamics of \(\rho_{11}\) and \(\rho_{LL}\) during the first pulse cycle. This observation extends through all subsequent pulse cycles. After the second cycle following pulses induce almost no more change in the population. Thus, the dynamics are quite different from the non-dephased case.

We also analyzed the emitted (or back-coupling) fields, by excluding any external source fields. Placing the emitter in the red position (Fig.~\ref{fig:circ}(e)), we take the initial condition \(\rho_{\text{RR}}=\rho_{11} = \frac{1}{2},\, \rho_{\text{LL}} = 0\)  with \(\rho_{1\text{R}}^{\text{Re}} = 0\) and \(\rho_{1\text{R}}^{\text{Im}} = \frac{1}{2}\). This results in an emitted field, which couples to the clockwise-propagating mode of the resonator. In Fig.~\ref{fig:circ}(f) we show a snapshot of \(S_0\) of the fields induced by the emitter backcoupling \(\bm{E}_{\text{bcpl}}\). After propagating along the resonator, the emitted field \(\bm{E}_{\text{bcpl}}\) couples to the bottom waveguide mode, propagating from left (\(x<0\)) to right (\(x>0\)), again confirming that our model can describe the directionality of the emission. \\

\section{\label{sec:fin}Conclusion}

In summary, we have developed a tool to analyze the dynamics of chiral coupling and non-reciprocal excitation occurring when coupling single emitters with nanophotonic structures. 

Specifically, we have developed a 2D-FDTD solver including the coupling to a V-type 3-level system with transitions sensitive to circularly polarised light (with opposite polarizations of the excited states). We then applied our model to a quantum emitter placed either in a dielectric waveguide or an optical circulator. We have shown that we are able to describe the polarization-dependent excitation of the emitter as well as the directionality of the emitted fields. 

For future research, our model can be readily extended to explore interactions with more complex structures like photonic crystals \cite{chiral_cpl_2dphc} or to include multiple emitters \cite{em_array_lattice}. Generally, in 2D simulations the induced backcoupling fields are small compared to the field strengths necessary to achieve a non-neglible excitation of additional emitters. Therefore, to couple many emitters and find stable solutions, a self-interaction free coupling approach like \cite{p_tfsf} would be advantageous. Having access to the time-resolved dynamics of nanophotonic systems coupled to quantum few-level systems will help developing novel devices in the field of quantum technology.

\begin{acknowledgments}
SSO acknowledges the support of the Sêr Cymru II Rising Star Fellowship (80762-CU145 (East)) which is part-funded by the European Regional Development Fund (ERDF) via the Welsh Government. 
\end{acknowledgments}

\appendix
\section{Optical Bloch equations for a 3-level V-type system}
The resulting Hamiltonian \(\cal{H_{\text{V}}}\) reads
\begin{eqnarray}
\arraycolsep=2.0pt\def\arraystretch{1.5}
 {\cal{H_{\text{V}}}} = \left( \begin{array}{ccc} 0 & -\bm{\mu}_{1R} \bm{E} & -\bm{\mu}_{1L} \bm{E} \\ 
 -\bm{\mu}_{1R}^* \bm{E}^* & \hbar \omega_R & 0\\
 -\bm{\mu}_{1L}^*t \bm{E}^* & 0 & \hbar \omega_L\\
\end{array}\right).
\end{eqnarray}
Applying Eq.\eqref{eq:lv} and including a dephasing rate \(\gamma\) (same rate for \(\rho_{1L}\) and \(\rho_{1R}\)) results in:
\begin{subequations}\begin{eqnarray}
i \hbar \frac{\partial}{\partial t}\rho_{11} &=&  -\rho_{1R}\bm{\mu}_{1R}^* \cdot \bm{E}^* -\rho_{1L}\bm{\mu}_{1L}^* \cdot \bm{E}^*\nonumber\\
& & +\rho_{1R}^*\bm{\mu}_{1R} \cdot \bm{E} +\rho_{1L}^*\bm{\mu}_{1L} \cdot \bm{E},\label{eq:rho11}\\
i\hbar\frac{\partial}{\partial t}\rho_{RR} &=& -\rho_{1R}^*\bm{\mu}_{1R} \cdot \bm{E} + \rho_{1R} \bm{\mu}_{1R}^* \cdot \bm{E}^*,\label{eq:rho22} \\
i\hbar\frac{\partial}{\partial t}\rho_{LL} &=& -\rho_{1L}^*\bm{\mu}_{1L} \cdot \bm{E} + \rho_{1L} \bm{\mu}_{1L}^* \cdot \bm{E}^*,\label{eq:rho33}\\
i\hbar\frac{\partial}{\partial t}\rho_{1R} &=& (\rho_{RR}-\rho_{11})\bm{\mu}_{1R} \cdot \bm{E} + \rho_{1R} \hbar \omega_R \nonumber\\
& & + \rho_{RL}^*\bm{\mu}_{1L} \cdot \bm{E} - i\hbar\gamma \rho_{1R},\label{eq:rho12}\\
i\hbar\frac{\partial}{\partial t}\rho_{1L} &=& (\rho_{LL}-\rho_{11})\bm{\mu}_{1L} \cdot \bm{E} + \rho_{1L} \hbar \omega_L \nonumber\\
& & + \rho_{RL}\bm{\mu}_{1R} \cdot \bm{E} - i\hbar\gamma \rho_{1L}, \label{eq:rho13}\\
i\hbar\frac{\partial}{\partial t}\rho_{RL} &=& -\rho_{1R}^*\bm{\mu}_{1L} \cdot \bm{E} + \rho_{RL} \hbar (\omega_L - \omega_R) \nonumber\\
& & + \rho_{1L} \bm{\mu}_{1R}^* \cdot \bm{E}^*. \label{eq:rho23}
\end{eqnarray}\end{subequations}
Using the same assumptions as in the 2-level emitter model, we can rewrite the dynamical equations to
\begin{subequations}\begin{eqnarray}
\frac{\partial}{\partial t}\rho_{11} &=& -\frac{2\mu}{\hbar}\left[ \rho_{1R}^{\text{Re}}(E_X^{\text{Re}} - E_Y^{\text{Im}}) + \rho_{1R}^{\text{Im}}(E_Y^{\text{Re}} + E_X^{\text{Im}}) \right]\nonumber\\
& & -\frac{2\mu}{\hbar}\left[ -\rho_{1L}^{\text{Re}}(E_Y^{\text{Im}} + E_X^{\text{Re}}) + \rho_{1L}^{\text{Im}}(E_Y^{\text{Re}} - E_X^{\text{Im}}) \right]\nonumber\\
\frac{\partial}{\partial t}\rho_{RR} &=& +\frac{2\mu}{\hbar}\left[ \rho_{1R}^{\text{Re}}(E_X^{\text{Re}} - E_Y^{\text{Im}}) + \rho_{1R}^{\text{Im}}(E_Y^{\text{Re}} + E_X^{\text{Im}}) \right]\nonumber\\
\frac{\partial}{\partial t}\rho_{LL} &=& +\frac{2\mu}{\hbar}\left[ -\rho_{1L}^{\text{Re}}(E_Y^{\text{Im}} + E_X^{\text{Re}}) + \rho_{1L}^{\text{Im}}(E_Y^{\text{Re}} - E_X^{\text{Im}}) \right]\nonumber
\end{eqnarray}\end{subequations}
\begin{subequations}\begin{eqnarray}
\frac{\partial}{\partial t}\rho_{1R} &=&
- i\frac{\mu}{\hbar} [ (\rho_{RR} - \rho_{11}) (E_Y^{\text{Re}} + E_X^{\text{Im}})\nonumber \\& & + \rho_{RL}^{\text{Re}}(E_Y^{\text{Re}} - E_X^{\text{Im}}) + \rho_{RL}^{\text{Im}}(E_Y^{\text{Im}} + E_X^{\text{Re}})]\nonumber\\
& & - i[\rho_{1R}^{\text{Re}}\omega_R + \rho_{1R}^{\text{Im}}\gamma]\nonumber\\
& & +\frac{\mu}{\hbar} [ (\rho_{RR} - \rho_{11}) (E_Y^{\text{Im}} - E_X^{\text{Re}})\nonumber \\& & + \rho_{RL}^{\text{Re}}(E_Y^{\text{Im}} + E_X^{\text{Re}}) - \rho_{RL}^{\text{Im}}(E_Y^{\text{Re}} - E_X^{\text{Im}})]\nonumber\\
& & + \rho_{1R}^{\text{Im}}\omega_R - \rho_{1R}^{\text{Re}}\gamma\nonumber\\
\frac{\partial}{\partial t}\rho_{1L} &=&
- i\frac{\mu}{\hbar} [ (\rho_{LL} - \rho_{11}) (E_Y^{\text{Re}} - E_X^{\text{Im}})\nonumber \\& & + \rho_{RL}^{\text{Re}}(E_Y^{\text{Re}} + E_X^{\text{Im}}) + \rho_{RL}^{\text{Im}}(E_X^{\text{Re}} - E_Y^{\text{Im}})]\nonumber\\
& & -i[\rho_{1L}^{\text{Re}}\omega_L + \rho_{1L}^{\text{Im}}\gamma]\nonumber\\
& & +\frac{\mu}{\hbar} [ (\rho_{LL} - \rho_{11}) (E_Y^{\text{Im}} + E_X^{\text{Re}})\nonumber \\& & + \rho_{RL}^{\text{Re}}(E_Y^{\text{Im}} - E_X^{\text{Re}}) - \rho_{RL}^{\text{Im}}(E_Y^{\text{Re}} + E_X^{\text{Im}})]\nonumber\\
& & + \rho_{1L}^{\text{Im}}\omega_L - \rho_{1L}^{\text{Re}}\gamma\nonumber\\
\frac{\partial}{\partial t}\rho_{RL} &=& - i\frac{\mu}{\hbar} [ \rho_{1L}^{\text{Re}}(E_Y^{\text{Re}} + E_X^{\text{Im}}) +\rho_{1L}^{\text{Im}}(E_Y^{\text{Im}} - E_X^{\text{Re}})\nonumber \\& & -\rho_{1R}^{\text{Re}}(E_Y^{\text{Re}} - E_X^{\text{Im}}) - \rho_{1R}^{\text{Im}}(E_Y^{\text{Im}} + E_X^{\text{Re}})]\nonumber\\
& & -i \rho_{RL}^{\text{Re}} (\omega_3 - \omega_2)\nonumber\\
& & +\frac{\mu}{\hbar} [\rho_{1L}^{\text{Re}} (E_X^{\text{Re}} - E_Y^{\text{Im}}) + \rho_{1L}^{\text{Im}} (E_X^{\text{Im}} + E_Y^{\text{Re}})\nonumber\\ & & - \rho_{1R}^{\text{Re}} (E_Y^{\text{Im}} + E_X^{\text{Re}}) - \rho_{1R}^{\text{Im}} (E_X^{\text{Im}} - E_Y^{\text{Re}})]\nonumber\\ & & + \rho_{RL}^{\text{Im}} (\omega_3 - \omega_2)\nonumber
\end{eqnarray}\end{subequations}

While the emitter in most FDTD implementations is coupled only to a single field component, this will hinder the description of chiral effects. Instead the quantum emitter should have a unified population for the interaction with all field components in the update equations as it is done in the  implementation presented here.

\providecommand{\noopsort}[1]{}\providecommand{\singleletter}[1]{#1}%

\end{document}